\documentclass[onecolumn,prb,aps,amsmath,amssymb,superscriptaddress,showpacs]{revtex4-2}
\usepackage{amsmath,mathtools}

\newcommand\nindent{.5pt}
\newcommand\noverline[1]{%
  \kern\nindent\overline{\kern-\nindent#1\kern-\nindent}\kern\nindent}

\usepackage{scalerel}
\usepackage[only,llbracket,rrbracket,llparenthesis,rrparenthesis]{stmaryrd}
\usepackage{accsupp}
\newcommand*{\llbrace}{%
  \BeginAccSupp{method=hex,unicode,ActualText=2983}%
    \textnormal{\usefont{OMS}{lmr}{m}{n}\char102}%
    \mathchoice{\mkern-4.05mu}{\mkern-4.05mu}{\mkern-4.3mu}{\mkern-4.8mu}%
    \textnormal{\usefont{OMS}{lmr}{m}{n}\char106}%
  \EndAccSupp{}%
}
\newcommand*{\rrbrace}{%
  \BeginAccSupp{method=hex,unicode,ActualText=2984}%
    \textnormal{\usefont{OMS}{lmr}{m}{n}\char106}%
    \mathchoice{\mkern-4.05mu}{\mkern-4.05mu}{\mkern-4.3mu}{\mkern-4.8mu}%
    \textnormal{\usefont{OMS}{lmr}{m}{n}\char103}%
  \EndAccSupp{}%
}

\newcommand{\threej}[6]{\left( \begin{array}{@{}c@{\;}c@{\;}c@{}}
         #1 & #2 & #3 \\
         #4 & #5 & #6
         \end{array}\right)}
\newcommand{\sixj}[6]{\left\{ \begin{array}{@{}c@{\;}c@{\;}c@{}}
         #1 & #2 & #3 \\
         #4 & #5 & #6
         \end{array}\right\}}

\usepackage{xfrac}

\usepackage{cancel}
\usepackage[english]{babel}

\usepackage{amsmath}
\usepackage{graphicx}
\usepackage[version=4]{mhchem}
\usepackage{bm}
\usepackage{dcolumn}
\usepackage{diagbox}

\begin{document}

\title{General expressions for Stevens and Racah operator equivalents}
\author{O.~Duros}\email{octave.duros@sorbonne-universite.fr}
\affiliation{Sorbonne Université, CNRS, Laboratoire de Chimie Physique - Matière et Rayonnement, LCP-MR, F-75005 Paris, France}%
\affiliation{Sorbonne Université, Muséum National d'Histoire Naturelle, CNRS, Institut de Minéralogie, de Physique des Matériaux et de Cosmochimie, IMPMC, F-75005 Paris, France}%
\author{A.~Juhin}
\affiliation{Sorbonne Université, Muséum National d'Histoire Naturelle, CNRS, Institut de Minéralogie, de Physique des Matériaux et de Cosmochimie, IMPMC, F-75005 Paris, France}%
\author{H.~Elnaggar}
\affiliation{Sorbonne Université, Muséum National d'Histoire Naturelle, CNRS, Institut de Minéralogie, de Physique des Matériaux et de Cosmochimie, IMPMC, F-75005 Paris, France}%
\author{G.~S.~Chiuzb\u aian}
\affiliation{Sorbonne Université, CNRS, Laboratoire de Chimie Physique - Matière et Rayonnement, LCP-MR, F-75005 Paris, France}%
\author{C.~Brouder}
\affiliation{Sorbonne Université, Muséum National d'Histoire Naturelle, CNRS, Institut de Minéralogie, de Physique des Matériaux et de Cosmochimie, IMPMC, F-75005 Paris, France}%

\date{\today}

\begin{abstract}
Several definitions of the crystal field have been used over time and their variety has lead to many misunderstandings, in both theoretical and experimental literature. Two categories of definitions can be mentioned, the first being the operators equivalents introduced by Stevens in 1952 and the second being the crystal-field operators, introduced by different authors from 1962 and expanded on the Racah spherical tensors. This paper aims at providing some clarification in this field. We first make a review of several expressions introduced in various references to compute crystal-field operators and we describe connections between them. Then, we introduce an explicit way to compute crystal-field operators, in terms of angular momentum quantum numbers $j$ and $m$ as well as in terms of $J^2$ and $J_z$ operators. We eventually give some connections between the Stevens operators equivalents and the crystal-field operators, and make usage of the coefficients of fractional parentage for the expression of the crystal-field operators for the many-body states. Various computational codes, using different crystal-field conventions, are also reviewed.
\end{abstract}

\maketitle

\section{Introduction}
\label{sec:intro}
When studying rare-earth ions, usually encountered as trivalent cations $R^{3+}$ of valence shell $\ell = f$, and transition metals ions of valence shell $\ell = d$, the first part of the Hamiltonian that one has to consider is the free-ion Hamiltonian, including spin-orbit coupling. Each energy level of the $\ell^n$ configuration is then $(2J+1)$-fold degenerate. By considering the ion in a crystal structure, the spherical symmetry is broken and each degenerate level splits under the influence of the electric field produced by the environment, also known as the crystal field (CF). As the crystal-field interactions in transition metals have already been extensively studied in the literature, the paper describes the more complex case of crystal field in rare-earths systems, even though the description is applicable to any metal.
The 4\textit{f} shell of most rare earths can be described by a spectroscopic term ${}^{2S+1}L_J$, where $J$ is a good quantum number. As a result, its ground state belongs to the vector space generated by state vectors $|JM\rangle$ (where $M=-J,\dots,J$). One has thus to consider, in the total Hamiltonian, the potential provided by the environment, whose symmetry depends on the point-group symmetry of the rare-earth ion. This potential, or its associated crystal-field Hamiltonian, has been subject to many different descriptions from several authors with almost as many different motivations.
From its definition as a perturbation of the electron cloud of the central ion by all the electrons of the system, the crystal-field Hamiltonian $H_\text{CF}$ is defined by:

\begin{equation*}
    H_\text{CF} = -e\sum_{i=1}^nV(r_i)
\end{equation*}
where $V(r_i)$ is the potential felt by the electron $i$ at its position $r_i$, $e$ is the elementary charge and the summation runs over all the $n$ electrons $i$ of the central ion. The simplest form of $V(r_i)$ is

\begin{equation*}
    V(r_i)=\sum_\text{L} \frac{(-Ze)_\text{L}}{|R_\text{L}-r_i|},
\end{equation*}
making the approximation this potential is created by discrete point charges $(-Ze)_\text{L}$ at positions $R_\text{L}$, where $Z$ is the formal charge of the surrounding ligands L (e.g. $Z=2$ for \ce{O^{2}-}). This resulted in the first point-charge models (PCM) description of the crystal field: by assuming the charges to be localized on positions $R_\text{L}$, the potential satisfies Laplace equation, with $\Delta V(r_i) = 0$. $V(r_i)$ can then be expressed as a sum of products of spherical harmonics times the crystal-field parameters $A^k_q r^k$~\cite{gorller-walrand_chapter_1996}:

\begin{equation}
   V(r_i) =\sum_{kq} A^k_q r^k Y_{kq}(\theta,\phi)
   \label{PCMpotential}
\end{equation}
where the parameters $A^k_q r^k$ can be fitted in an empirical way. This assumption is actually wrong, because the charge density is delocalized. Hence $\Delta V \ne 0$ and any arbitrary potential can be written as a sum of products of spherical harmonics times a radial function $f_{kq}(\hat{r})$, as~\cite{kalf_expansion_1995}:

\begin{equation}
   V(r_i) =\sum_{kq} f_{kq}(\hat{r}) Y_{kq}(\theta,\phi).
   \label{KalfPotential}
\end{equation}
It is in this general framework that we shall now continue. Operator equivalents were introduced in 1952 by Stevens~\cite{stevens_matrix_1952} to parameterize this potential and the action of the crystal field was written as a linear combination of operators that can be expressed as a polynomial in the total angular momentum operators $J_x$, $J_y$ and $J_z$. The associated Hamiltonian is then defined like the PCM potential in Eq. (\ref{PCMpotential}), but with a sum over tensor operators acting on the total angular momentum operators rather than over the spherical harmonic acting on the positions of the ligands. In his first definition, Stevens did not give a proper name to these operator equivalents. He only presented his way to replace the Cartesian $x$, $y$, $z$ coordinates with the total angular momentum operators $\mathbf{J}$ in the tesseral harmonic functions, taking the prefactors out at the same time. His operators equivalents were later denoted $O_n^m$ by Baker \textit{et al.}~\cite{baker_paramagnetic_1958} and $O_k^q$ by Abragam and Bleaney~\cite{abragam_electron_1970}(where $n$ is analogous to the rank $k$, and $m$ is analogous to the quantum magnetic number $q=-k,\dots,k$).
These Stevens operator equivalents, also shortened Stevens operators, were then widely used to analyze spectroscopic experiments in electron paramagnetic resonance (EPR), electron spin resonance (ESR), electron nuclear double resonance (ENDOR), UV-visible spectroscopy,  M\"ossbauer spectroscopy~\cite{rudowicz_generalization_2004} as well as neutron scattering and x-ray absorption spectroscopy. The expression of operator equivalents were constructed according to needs: Stevens firstly introduced $O_2^0$, $O_4^0$, $O_6^0$ and $O_6^6$, then, $O_6^6+O_6^{-6}$ was given by Bleaney and Stevens~\cite{bleaney_paramagnetic_1953}. As a consequence, the transformation properties of $O_k^q$ under rotation are very complicated and the operators of different $q$ have no simple relations. The absence of an analytical expression for these operator equivalents imposed the publication of extensive tables trying to provide explicit expressions for $O_k^q$. Running over many pages and scattered in several articles and books, each new table lists the misprints of the previous ones. Moreover, a single entry can be quite complicated (for example the expression for $O_{12}^1$ is 11 lines long in~\cite{rudowicz_generalization_2004}).
It was later realized~\cite{koster_method_1959} that, by contrast, spherical tensor operators $T^{(k)}_q$ (of rank $k$ and with $-k<q<k$)~\cite{racah_theory_1942} undergo the same transformation properties as spherical harmonics and are linked by simple relations (see sec.~\ref{sec:def&main_prop}). 
In particular, spherical tensors can be used to express a Hamiltonian with any symmetry. 
For all these reasons, many authors, like Wybourne~\cite{wybourne_spectroscopic_1965} replaced Stevens operators by spherical tensors operators or combinations thereof, leading to new tables. Besides, additional confusion arose from the fact that spherical operators tensors come with various normalization in the literature. Rudowicz \textit{et al.} reviewed these tables in two publications~\cite{rudowicz_transformation_1985,rudowicz_generalization_2004}.

To avoid egregious mistakes, some authors proposed to derive these results with a computer through recursion methods~\cite{bose_operator_1975,ryabov_generation_1999}.
Several closed form expressions were also proposed~\cite{caola_operator_1974} to solve the aforementioned problems. The plan of the paper is the following. We start from the definition of spherical tensors operators $T^{(k)}_q$. We then introduce the operator $P^{(k)}_q$ which is a polynom in $j$ and $m$ and that has the advantage of having only diagonal matrix elements ($\langle jm| P^{(k)}_q|jm'\rangle = 0$ if $m=m'$), from which $T^{(k)}_q$ can be easily computed. Then, we review the different expressions for $P^{(k)}_q$ proposed in the literature and we illustrate connections between them. We also present a new method to relate the expression of $\langle jm| P^{(k)}_q|jm\rangle$ in terms of $j$ and $m$ to an analytical expression of $P^{(k)}_q$ in terms of $J^2$ and $J_z$. Finally, we connect various definitions and describe the way to define properly the crystal-field Hamiltonian with Racah's unit tensor operator, based on the coefficients of fractional parentage (cfp).


     

\section{Spherical tensors and their matrix elements}
In continuity with the work of Condon and Shortley~\cite{condon_theory_1935}, Racah~\cite{racah_theory_1942, racah_theory_1943, racah_theory_1949} developed the method of tensor operator $\mathbf{T}^{(k)}$ with components $T^{(k)}_q$ to calculate matrix elements of the interactions between two $\mathit{\tau LSJM}$ states. $\mathit{L}$ defines the orbital angular momentum operator, $\mathit{S}$ the spin operator, $J$ the total angular momentum operator, $\mathit{M} \equiv \mathit{M}_\mathit{J} = \mathit{M}_\mathit{L} + \mathit{M}_\mathit{S}$ is the value associated to $\mathit{J}_z$ operator and $\tau$ is all the others quantum numbers that must be used to differentiate states of same $\mathit{LSJM}$. In comparison to the determinant methods of Slater, this powerful method allowed the calculation of the complex electronic properties encountered in rare earths. Any Hamiltonian can thus be written as a sum:

\begin{align}
H &= \sum_{kq} B^k_q T^{(k)}_q.
\label{generalHamiltonian}
\end{align}

\subsection{Definition and main properties}
\label{sec:def&main_prop}
The irreducible tensor operator $\mathbf{T}^k$ of rank $k$ has $2k+1$ spherical tensors components $T^{(k)}_q$, where $q=-k,-k+1,\dots,k$. These components, or operators, satisfy the same commutation rules as the spherical harmonic operators $Y_{kq}$ with respect to angular momentum $J$ and are defined following the relations~\cite[p.~90, p.~157]{biedenharn_angular_1981}

\begin{align}
T^{(k)}_k &= a_k J_+^k,\label{Tkk}\\
\left[J_\pm,T^{(k)}_q\right] &= \sqrt{(k\mp q)(k\pm q+1)}T^{(k)}_{q\pm1},
\label{recurTkmu}
\end{align}
where $a_k$ is a real constant, $J_\pm=J_x \pm i J_y$ and $J_+^k=(J_+)^k$ (i.e. $k$ is a true exponent and not an upper index as in $T^{(k)}_q$). The notation $[a,b]$ corresponds to the commutation of $a$ and $b$. The $T^{(k)}_q$ operators satisfy the important properties~\cite[p.~656]{biedenharn_angular_1981}

\begin{align}
(T^{(k)}_q)^\dagger &= (-1)^q T^{(k)}_{-q},
\label{duality}
\end{align}
and

\begin{align}
\sum_{q=-k}^k (T^{(k)}_q)^\dagger T^{(k)}_q &= (a_k)^2 \frac{(k!)^2}{(2k)!}
\prod_{s=1}^k (4 J^2 +1 -s^2),
\label{TkmuTkmu}
\end{align}
where $J^2=J_x^2 + J_y^2 + J_z^2$.

Moreover, if for a given $j$,  we denote by $\text{M}^{(k)}_q$ the $(2j+1)\times(2j+1)$
matrix defined by the matrix elements
$(\text{M}^{(k)}_q)_{mm'}=  \langle jm|{T}^{(k)}_{q}|jm'\rangle$,
then we have the orthogonality relation~\cite{grenet_operator_1978}:

\begin{align}
\mathrm{Tr}\big( (\text{M}^{(k)}_q)^\dagger \text{M}^{(k')}_{q'}\big) &= \delta_{k,k'} \delta_{q,q'}\,
 \frac{(k!)^2 a_k^2}{(2k+1)!} \frac{(2j+k+1)!}{(2j-k)!},
\label{orthogonality}
\end{align}
where the adjoint $\dagger$ can be replaced by a transpose since the matrix elements are real.
This property can be used for instance to determine the expansion of a Hamiltonian over spherical tensors. If $H$ is a $(2j+1)$$\times$$(2j+1)$ Hamiltonian matrix, as given in Eq.~(\ref{generalHamiltonian}), then its parameters
$B^k_q$ introduced in Eq.~(\ref{generalHamiltonian}) are

\begin{align*}
B^k_q &=    \frac{(2k+1)!}{ (k!)^2 a_k^2} \frac{(2j-k)!}{(2j+k+1)!}
\mathrm{Tr}\big( (\text{M}^{(k)}_q)^\dagger H    \big).
\end{align*}

\subsection{Common conventions}

We denote by $T^{(k)}_q$ the spherical tensor operator from Eq.~(\ref{Tkk}) corresponding to~\cite[p.~656]{biedenharn_angular_1981}

\begin{align*}
 a_k &= (-1)^k \frac{\sqrt{(2k)!}}{k!}.
\end{align*}
$T^{(k)}_q$ is actually note unique and is part of a family defined by different $a_k$. For instance, another frequently used convention~\cite{tuszynski_spherical_1990} comes with

\begin{align*}
 a_k &= (-1)^k 2^{-\sfrac{k}{2}}.
\end{align*}

On his side, Racah~\cite{racah_theory_1942} defined the unit tensor operators $u^{(k)}_q(j)$ with $a_k$ dependent on
the value of $j$ on which the tensor operators are applied:

\begin{align}
\label{eq:a_kRacah}
a_k &=
 \frac{(-1)^k}{k!}
\sqrt{\frac{(2k)! (2j-k)!}{(2j+k+1)! }}.
\end{align}

Let us finally define the spherical tensor operators ${C}^{(k)}_q$ used by Wybourne~\cite{wybourne_spectroscopic_1965} with:
\begin{align}
    \label{eq:a_kWybourne}
    a_k &= (-1)^{2j-\sfrac{k}{2}}(2j+1)\left[\frac{(j+\frac{k}{2})!}{(\frac{k}{2})!(\frac{k}{2})!\big(j-\frac{k}{2}\big)!}\right]^{\sfrac{1}{2}}\frac{(2j-k)!}{(2j+k+1)!}\sqrt{(2k)!}.
\end{align}

\subsection{Matrix elements}

The matrix elements of the Hamiltonian in Eq.~(\ref{generalHamiltonian}) can be derived
from the matrix elements of $T^{(k)}_q$, which are
given by the Wigner-Eckart theorem:

\begin{align*}
\langle j m'| T^{(k)}_q|jm\rangle &=
(-1)^{j-m'}  \threej{j}{k}{j}{-m'}{q}{m} \langle  j\|T^{(k)}\|j\rangle .
\end{align*}
To compute the reduced matrix element $\langle  j\|T^{(k)}\|j\rangle$,
one only needs

\begin{align}
T^{(k)}_k |jm\rangle &= a_k J_+^k |jm\rangle
&= a_k \left(\frac{(j-m)!(j+m+k)!}{(j-m-k)!(j+m)!}\right)^{\sfrac{1}{2}} |j m+k\rangle,
\label{Jpluskjm}
\end{align}
and the explicit expression~\cite[p.~268]{khersonskii_quantum_1988}

\begin{align*}
\threej{j}{k}{j}{-m'}{k}{m} &= \delta_{m',m+k} (-1)^{j-m}
\left(\frac{(2k)! (2j-k)! (j+m+k)!(j-m)!}{(2j+k+1)! (k!)^2 (j-m-k)!(j+m)!}\right)^{\sfrac{1}{2}},
\end{align*}
which gives

\begin{align}
\langle  j\|T^{(k)}\|j\rangle &=
 (-1)^k a_k k!
\left(\frac{(2j+k+1)! }{(2k)! (2j-k)!}\right)^{\sfrac{1}{2}}.
\label{reducedME}
\end{align}
In particular, it is seen that with $a_k$ from Eq.~(\ref{eq:a_kRacah}), $\langle  j\|T^{(k)}\|j\rangle=1$, hence the name chosen by Racah for his unit tensor $\textbf{u}^{(k)}$. Note that Eq.~\eqref{reducedME} differs from the
reduced matrix element given in~\cite[p.~157]{biedenharn_angular_1981} by a factor of $\sqrt{2j+1}$ due to the fact that these authors use a Wigner-Eckart theorem involving Clebsch-Gordan
coefficients instead of 3-$j$ symbols. Note that the orthogonality relation~\eqref{orthogonality} follows from the Wigner-Eckart theorem and the orthogonality of 3-$j$ symbols. 


     

\section{Explicit expressions for $T^{(k)}_q$}
In this section we assume that $q\ge0$.
The first step to obtain a explicit expression is to write $T^{(k)}_q$
as the product of $J_+^q$ times a polynomial in $J^2$ and $J_z$, introduced here as $P^{(k)}_q$.
This polynomial will only have non-zero matrix elements between vector states with
the same $m$ and $j$.   
The second step is to obtain explicit expressions for this polynomial.

\subsection{$P^{(k)}_q$ and its properties}
To derive an explicit expression for $T^{(k)}_q$, we first define $P^{(k)}_q$ for $q\ge0$ by~\cite[p.~658]{biedenharn_angular_1981}  

\begin{align}
T^{(k)}_q &= a_k \sqrt{\frac{(k+q)!(k-q)!}{(2k)!}}
J_+^q P^{(k)}_q,
\label{TkmuJpmuPkmu}
\end{align}
with $P^{(k)}_k=1$.
We first show that $P^{(k)}_q$ is a polynomial in $J^2$ and $J_z$.
Indeed, the recursive relation~\eqref{recurTkmu} for $T^{(k)}_q$ leads to
the recursive expression for $P^{(k)}_q$

\begin{align*}
(k-q+1) P^{(k)}_{q-1} &= P^{(q)}_{q-1} P^{(k)}_q
+
J_+\left[J_-,P^{(k)}_q\right],
\end{align*}
where

\begin{align*}
P^{(q)}_{q-1} &=  -q(q-1) - 2 q  J_z.
\end{align*}
To give a recursive proof of the polynomial nature of
$P_q$, we start from the fact that $P^{(k)}_k=1$ is a polynomial in $J^2$ and $J_z$.
Assume now that $P^{(k)}_q$ is a polynomial in $J^2$ and $J_z$. Then,
by using~\cite{bose_operator_1975}

\begin{align*}
\left[J_-,J_z^n\right] &= 
J_-\nabla J_z^n,
\end{align*}
where $\nabla J_z^n=J_z^n -(J_z-1)^n$ is the backward difference, we see that 
$\left[J_-,P^{(k)}_q\right]$ is the product of $J_-$ by a polynomial in
$J^2$ and $J_z$. It follows that $J_+\left[J_-,P^{(k)}_q\right]$ and 
$P^{(k)}_{q-1}$ are also polynomials of this form
because $J_+J_-=J^2 - J_z^2-J_z$.
As a consequence, the operator $P^{(k)}_q$ is a polynomial in $J^2$ and $J_z$
for all $q$ such that $k\ge q\ge0$, which implies that its matrix elements
are diagonal in $j$ and $m$:
$\langle jm|P^{(k)}_q|j'm'\rangle=\delta_{jj'}\delta_{mm'}
\langle jm|P^{(k)}_q|jm\rangle$ .

To be more explicit, if $P^{(k)}_q(J^2,J_z)$ is the operator $P^{(k)}_q$ whose dependence
on $J^2$ and $J_z$ is made explicit, we denote by 
$P^{(k)}_q(J^2,\nabla J_z)$ the operator $P^{(k)}_q(J^2,J_z)$
where each monomial $ J_z^n$ is replaced by $\nabla J_z^n$.
Then, the recursive expression becomes

\begin{align*}
(k-q+1) P^{(k)}_{q-1}(J^2,J_z) &=  -\big(q(q-1) + 2 q  J_z\big) P^{(k)}_q(J^2,J_z) 
+ (J^2 - J_z^2-J_z)P^{(k)}_q(J^2,\nabla J_z).
\end{align*}

Finally, we give the formula for $T^{(k)}_{-q} $ 
following Ryabov~\cite{ryabov_generation_1999}

\begin{align*}
T^{(k)}_{-q} &= (-1)^k
a_k
\sqrt{\frac{(k+q)!(k-q)!}{(2k)!}}
J_-^q P^{(k)}_q(J^2,-J_z).
\end{align*}
If we combine this with $(T^{(k)}_{q})^\dagger=(-1)^qT^{(k)}_{-q}$,
$J_+^\dagger=J_-$ and the fact that $P^{(k)}_q(J^2,J_z)$ is self-adjoint,
we obtain the remarkably simple intertwining formula

\begin{align*}
 P^{(k)}_q(J^2,J_z) J_-^q &= (-1)^{k-q}
J_-^q P^{(k)}_q(J^2,-J_z).
\end{align*}
We now list the existing closed-form expressions for the matrix elements of $P^{(k)}_q$. Note that each expression corresponds to a specific formula for the corresponding Clebsch-Gordan coefficients. There is a very large number of expressions for Clebsch-Gordan coefficients, but only very few are suited to our purpose. A large number of such formulas are given
in~\cite{khersonskii_quantum_1988} and Shimpuku gives 168 different
formulas in~\cite{shimpuku_expressions_1963}. Below, we list the most useful ones.

\subsection{Caola}
The first closed-form expression for $P^{(k)}_{q}$ was given by Caola~\cite{caola_operator_1974}

\begin{align*}
\langle jm'| T ^{(k)}_q| jm\rangle &=
(-1)^k \Big(\frac{(2j+1)(k!)^2 (2j-k)!}{(2j+k+1)!}\Big)^{\sfrac{1}{2}}
\sqrt{(k+q)!(k-q)!} 
\\\times &\sum_{p=q}^k (-1)^p 
\frac{(j-q-m)! (j+m)! }{(j-m-p)!(j-k+m+p)! p! (k-p)!(p-q)!(k+q-p)!}
\langle jm'| J_+^q |jm\rangle,
\end{align*}
where we corrected a misprint.
By comparing this expression with Eq.~(3.170) in~\cite[p.~79]{biedenharn_angular_1981} for the Clebsch-Gordan coefficients $C^{jm'}_{jmkq}$ we see that

\begin{align*}
\langle jm'| T^{(k)}_q|jm\rangle &=
C^{jm'}_{jmkq},
\end{align*}
so that

\begin{align*}
a_k &=
 \frac{(-1)^k}{k!}
\left(\frac{(2j+1)(2k)! (2j-k)!}{(2j+k+1)! }\right)^{\sfrac{1}{2}}.
\end{align*}
Therefore,

\begin{align*}
\langle jm| P^{(k)}_q|jm\rangle &=
(-1)^k (k!)^2\sum_{p=q}^k (-1)^p
\frac{(j-q-m)! (j+m)! }{(j-m-p)!(j-k+m+p)! p! (k-p)!(p-q)!(k+q-p)!}.
\end{align*}
By using the descending Pochhammer symbol
 $\left[a\right]_p$ defined for $p\ge0$ by
 
\begin{align}
\left[a\right]_p &= 
\frac{a!}{(a-p)!},
 \end{align}
we can rewrite this as

\begin{align}
\langle jm| P^{(k)}_q|jm\rangle &=
(-1)^k \sum_{p=q}^k (-1)^p
\binom{k}{p}\binom{k}{p-q}
{[}j+m{]}_{k-p} {[}j-m-q{]}_{p-q}.
\label{PkmnuCaola}
\end{align}

\subsection{Biedenharn and Louck}
Table 6 of \cite{biedenharn_angular_1981} gives the following relation for the non-zero matrix elements of $P^{(k)}_q$ :

\begin{align}
\langle j m| P^{(k)}_q | jm\rangle &=  
\sum_{p=0}^{k-q} (-1)^p \binom{k}{p}\binom{k}{q+p} \left[j+m\right]_p
\left[j-m-q\right]_{k-q-p}.
\label{Pkmujm}
 \end{align}
where $\left[a\right]_p$ is the descending Pochhammer symbol for $p\ge0$

\begin{align}
\left[a\right]_p &=  
\frac{a!}{(a-p)!}.
 \end{align}
Note that in the rest of their book, Biedenharn and Louck use the ascending Pochhammer symbol~\cite[p.~352]{biedenharn_angular_1981}. We notice that the expressions given by Caola and by Biedenharn and Louck become identical if we write $q=k-p$.

\subsection{Grenet and Kibler}
Again for $q\ge0$, the formula given by 
Grenet and Kibler is~\cite{grenet_operator_1978}

\begin{align*}
 \langle jm| P^{(k)}_q|jm\rangle &=  \frac{k!}{(2j-k)!(k+q)!}
 \sum_z (-1)^{q+z} \frac{(2j-z)!(k+z)!}{z!(k-z)!(z-q)!}
\frac{(j+m)!}{(j+m+q-z)!},
\end{align*}
that we rewrite

\begin{align}
 \langle jm| P^{(k)}_q|jm\rangle &=  \frac{(k!)^2}{(k+q)!}
 \sum_{p=0}^{k-q} \frac{(-1)^{p}}{p!} 
\binom{k+q+p}{k}\binom{2j-q-p}{2j-k}\left[j+m\right]_p.
\label{GKformula}
\end{align}
A remarkable aspect of this formula is that $m$ appears only in 
one term $\left[j+m\right]_p$. 
This allows us to obtain a relatively simple expression for
$P^{(k)}_q$ as a function of $J_z$. Indeed, if we expand the Pochhammer 
symbol in terms of (signed)
Stirling numbers of the first kind $s(p,n)$:

\begin{align}
\left[a\right]_p &=  
\sum_{q=0}^p s(p,q) a^q.
 \end{align}
Hence

\begin{align}
\left[j+m\right]_p &=  
\sum_{q=0}^p s(p,q) (j+m)^q=
\sum_{n=0}^p m^n \sum_{q=0}^{p-n} \binom{n+q}{n} s(p,n+q) j^q
 \end{align}

And we obtain an explicit expression of $P^{(k)}_q$ in terms of $j$ and $J_z$:

\begin{align}
P^{(k)}_q &=  
\sum_{n=0}^{k-q} \alpha^{k}_{q n}(j) J^n_z,
 \end{align}
where

\begin{align}
\alpha^{k}_{q n}(j) &= 
\frac{(k!)^2}{(k+q)!}
 \sum_{p=n}^{k-q} \frac{(-1)^{p}}{p!} 
\binom{k+q+p}{k}\binom{2j-q-p}{2j-k}
\sum_{q=0}^{p-n} \binom{n+q}{n} s(p,n+q) j^q.
\end{align}

\subsection{Other expressions}
Since matrix elements of $T^{(k)}_q$ are directly related to Clebsch-Gordan coefficients, every Clebsch-Gordan coefficient formula gives an expression for the matrix elements of $T^{(k)}_q$. 
The formulas used in the literature to calculate
$T^{(k)}_q$ correspond to the
cases where there are a small number of terms using
the variable $m$, so that the expression in terms of $J_z$
is (relatively) simple. From that point of view, the formula given by Grenet and Kibler cannot be improved.

There are also several semi-explicit expressions
using iterated forward differences, quasi-powers or hypergeometric functions~\cite{khersonskii_quantum_1988}.
Within this semi-explicit family, 
a series of authors~\cite{meckler_majorana_1958, meckler_algebra_1959, normand_relations_1982, marinelli_simplified_1983, siminovitch_very_2022} used
a connection between some Clebsch-Gordan coefficients and Chebyshev polynomials of a discrete real variable (not to be confused with the Chebyshev polynomials $T_n(x)$ and $U_n(x)$ of the first and second kind). 
Then, they used iterated forward differences of these polynomials to obtain the Clebsch-Gordan coefficients required to calculate the matrix elements of $T^{(k)}_q$. We do not elaborate because semi-explicit formulas are not the subject of this paper and because a very nice and extensive article was recently published on this topic~\cite{siminovitch_very_2022}.

\subsection{From matrix elements to operators}
We proved that $P^{(k)}_q$ is a polynomial of $J^2$ and $J_z$. We gave 
an explicit expression for $P^{(k)}_q$ as a polynomial of $J_z$ with
coefficients $\alpha^k_{q m}(j)$ which are polynomials of $j$. 
It is interesting to use expressions for $P^{(k)}_q$ as a polynomial of $J_z$
and $J^2$ because these expressions are independent of $j$ and some
tables are given in terms of $J^2$ instead of $j$.
To replace the dependence on $j$ by a
dependence on $J^2$, we need to find polynomials
$\beta^k_{q m}(J^2)$ such that 

\begin{align*}
\beta^k_{q m}(J^2)|jm\rangle &= \beta^k_{q m}\big(j(j+1)\big)|jm\rangle
=\alpha^k_{q m}(j)|jm\rangle.
\end{align*}

We treat this problem in a general way and for notational convenience
we replace $j$ by $x$ and we solve the following problem:
let $P$ and $Q$ be polynomials related by 
$P(x) = Q\big(x(x+1)\big) $, we calculate the coefficients of $Q$ from the
coefficients of $P$.
Assume that

\begin{align*}
P(x) &= \sum_{i=0}^n a_i x^i = Q\big(x(x+1)\big) = \sum_{j=0}^m b_j\big(x(x+1))^j.
\end{align*}
We expand

\begin{align*}
 Q\big(x(x+1)\big) &= \sum_{j=0}^m \sum_{k=0}^j b_j \binom{j}{k}
x^{2k}x^{j-k}
= \sum_{j=0}^m \sum_{k=0}^j b_j \binom{j}{k}x^{j+k}.
\end{align*}
In the last expression we define $i=j+k$, so that $k=i-j$. We know that
$j\ge k\ge0$ so that $i/2\le j\le i$. But we also know that $0\le j\le m$.
Thus 

\begin{align*}
 Q\big(x(x+1)\big) &= 
 \sum_{i=0}^{2m} x^i  \sum_{j=\lceil i/2\rceil}^{\min(m,i)} b_j \binom{j}{i-j},
\end{align*}
and 

\begin{align*}
a_i &=    \sum_{j=\lceil i/2\rceil}^{\min(m,i)} b_j \binom{j}{i-j}.
\end{align*}
The term of maximum degree is $a_{2m}=b_m$, so that $n=2m$.
If we want to expression $b_j$ in terms of $a_i$, we need only the
first $m+1$ values $a_i$. Moreover, we have $a_0=b_0$ and the 
interesting cases are

\begin{align*}
a_i &=   \sum_{j=1}^m M_{ij} b_j,
\end{align*}
where the $m\times m$ matrix $M=(M_{ij})$ defined by

\begin{align*}
M_{ij} &=     \binom{j}{i-j} \text{ for }i\ge j\ge \lceil i/2\rceil,\\
M_{ij} &= 0 \text{ otherwise}
\end{align*}
is invertible because it is lower diagonal with diagonal $M_{ii}=1$. 
Since $M$ is a lower triangular matrix with unit diagonal, its inverse is also a lower triangular matrix with unit diagonal.
To prove this, write $M=1-K$, where $K$ is lower triangular with zero diagonal 
so that $K^m=0$ and note that  $M^{-1}=I+\sum_{k=1}^{m-1}K^k$.
Moreover, the first $k$ lines of the inverse of a triangular matrix $M$  depend only on the
first $k$ lines of $M$. In this sense, the inverse of $M$ does not depend on the dimension of $M$.

It turns out that this problem is solved by Riordan's theorem~\cite[p.~50]{riordan_combinatorial_1979}:
\\if 

\begin{align*}
\alpha_i  &=    \sum_j (-1)^j \beta_j \binom{p+qj-j}{i-j},\\
\end{align*}
then

\begin{align*}
\beta_i  &=    \sum_j (-1)^j \alpha_j \frac{p+qj-j}{p+qi-j}\binom{p+qi-j}{i-j}.
\end{align*}
In the case $\alpha_i=a_i$, $\beta_j=(-1)^j b_j$, $p=0$ and $q=2$ we recover
our equation

\begin{align*}
a_i  &=    \sum_j  b_j \binom{j}{i-j},
\end{align*}
and Riordan's theorem gives the inverse relation:

\begin{align*}
b_0 &= a_0,\\
b_i &= \sum_{j=1}^{i} (-1)^{i-j}  \frac{j}{2i-j}\binom{2i-j}{i-j}a_j\text{ for } i\ge1.
\end{align*}

\subsection{Explicit expressions in terms of $j$ and $m$}
As an application of the formulae presented in this paper, we give
explicit expressions for $P^{(k)}_q$ in terms of $j$ and $m$
for $k=0$ to $6$.

\begin{align*}
P^{(0)}_0 &= 1.
\end{align*}
\begin{align*}
P^{(1)}_0 &= -2 m,\\
P^{(1)}_1 &= 1.
\end{align*}
\begin{align*}
P^{(2)}_0 &= -2 j^2-2 j+6 m^2,\\
P^{(2)}_1 &= -4 m-2,\\
P^{(2)}_2 &= 1.
\end{align*}
\begin{align*}
P^{(3)}_0 &= 12 j^2 m+12 j m-20 m^3-4 m,\\
P^{(3)}_1 &= -3 j^2-3 j+15 m^2+15 m+6,\\
P^{(3)}_2 &= -6 m-6,\\
P^{(3)}_3 &= 1.
\end{align*}
\begin{align*}
P^{(4)}_0 &= 6 j^4+12 j^3-60 j^2 m^2-6 j^2-60 j m^2-12 j+70 m^4+50 m^2,\\
P^{(4)}_1 &= 24 j^2 m+12 j^2+24 j
   m+12 j-56 m^3-84 m^2-76 m-24,\\
P^{(4)}_2 &= -4 j^2-4 j+28 m^2+56 m+36,\\
P^{(4)}_3 &=-8 m-12,\\
P^{(4)}_4 &=1.
\end{align*}
\begin{align*}
P^{(5)}_0 &= -60 j^4 m-120 j^3 m+280 j^2 m^3+140 j^2 m+280 j m^3+200 j m-252 m^5-420 m^3-48
   m,\\
P^{(5)}_1 &= 10 j^4+20 j^3-140 j^2 m^2-140 j^2 m-70 j^2-140 j m^2-140 j m-80 j+210 m^4 \\& +420
   m^3+630 m^2+420 m+120,\\
P^{(5)}_2 &= 40 j^2 m+40 j^2+40 j m+40 j-120 m^3-360 m^2-480 m-240,\\
P^{(5)}_3 &=-5 j^2-5
   j+45 m^2+135 m+120,\\
P^{(5)}_4 &=-10 m-20,\\
P^{(5)}_5 &=1.
\end{align*}
\begin{align*}
P^{(6)}_0 &= -20 j^6-60 j^5+420 j^4 m^2+100 j^4+840 j^3 m^2+300 j^3-1260 j^2 m^4-1680 j^2
   m^2-80 j^2\\&-1260 j m^4-2100 j m^2-240 j+924 m^6+2940 m^4+1176 m^2,\\
P^{(6)}_1 &= -120 j^4 m-60 j^4-240
   j^3 m-120 j^3+720 j^2 m^3+1080 j^2 m^2+1200 j^2 m+420 j^2+720 j m^3\\&+1080 j m^2+1320 j
   m+480 j-792 m^5-1980 m^4-4320 m^3-4500 m^2-2808 m-720,\\
P^{(6)}_2 &= 15 j^4+30 j^3-270 j^2 m^2-540
   j^2 m-375 j^2-270 j m^2-540 j m-390 j+495 m^4+1980 m^3\\&+4095 m^2+4230 m+1800,\\
P^{(6)}_3 &=60 j^2
   m+90 j^2+60 j m+90 j-220 m^3-990 m^2-1790 m-1200,\\
P^{(6)}_4 &= -6 j^2-6 j+66 m^2+264 m+300,\\
P^{(6)}_5 &=-12
   m-30,\\
P^{(6)}_6 &=1.
\end{align*}

\subsection{Explicit expressions in terms of $J^2$ and $J_z$}
As an application of the formulae presented in this paper, we give
explicit expressions for $P^{(k)}_q$ in terms of $J^2$ and $J_z$
for $k=0$ to $6$. 

\begin{align*}
P^{(0)}_0 &= 1.
\end{align*}
\begin{align*}
P^{(1)}_0 &= -2 J_z,\\
P^{(1)}_1 &= 1.
\end{align*}
\begin{align*}
P^{(2)}_0 &= -2 J^2+6 J_z^2,\\
P^{(2)}_1 &= -2( 2J_z+1),\\
P^{(2)}_2 &= 1.
\end{align*}
\begin{align*}
P^{(3)}_0 &=3J_z (4J^2 J_z- 5J_z^2-1),\\
P^{(3)}_1 &= -3 J^2+3( 5J_z^2 +5J_z+2),\\
P^{(3)}_2 &= -6 (J_z+1),\\
P^{(3)}_3 &= 1.
\end{align*}
\begin{align*}
P^{(4)}_0 &= 6 (J^2)^2-12 J^2 (5 J_z^2+1)+10
J_z^2(7 J_z^2+5),\\
P^{(4)}_1 &=3 (2J_z+1) (J^2 -7J_z^2-7 J_z-6),\\
P^{(4)}_2 &= -4 J^2+4(7 J_z^2+14 J_z+9),\\
P^{(4)}_3 &=-4(2J_z+3),\\
P^{(4)}_4 &=1.
\end{align*}
\begin{align*}
P^{(5)}_0 &= -4J_z\big(15 (J^2)^2+10J^2 (7J_z^2+5) -3(21J_z^4+35 J_z^2+4)\big),\\
P^{(5)}_1 &= 10 (J^2)^2-20J^2(7 J_z^2+7 J_z+4)
+30(7 J_z^4  +14
   J_z^3+21 J_z^2+14 J_z+4),\\
P^{(5)}_2 &=40 (J_z+1)\big(J^2 -3 (J_z^2+2J_z+2)\big),\\
P^{(5)}_3 &=-5 J^2
  +15 (3J_z^2+9 J_z+8),\\
P^{(5)}_4 &=-10 (J_z+2),\\
P^{(5)}_5 &=1.
\end{align*}
\begin{align*}
P^{(6)}_0 &= -20 (J^2)^3+20(J^2)^2(21 J_z^2+8)
-60J^2(21 J_z^4+35 J_z^2+4)
+84J_z^2(11J_z^4+35 J_z^2+14),\\
P^{(6)}_1 &= -12(2J_z+1)\big(5(J^2)^2
+10J^2(4 + 3 J_z + 3 J_z^2)-3
(11J_z^4+22J_z^3+49J_z^2+38J_z+20)\big),\\
P^{(6)}_2 &= 15(J^2)^2-30J^2(9J_z^2+18J_z+13)
+45 (11J_z^4+44 J_z^3+91 J_z^2+94 J_z+40),\\
P^{(6)}_3 &=10(2J_z+3)(3J^2-
11J_z^2-33J_z-40),\\
P^{(6)}_4 &= -6 J^2+6(11 J_z^2+44 J_z+50),\\
P^{(6)}_5 &=-6(2
   J_z+5),\\
P^{(6)}_6 &=1.
\end{align*}

We can notice that $P^{(2m)}_{2\nu+1}$ is proportional
to $2J_z+2\nu+1$ and $P^{(2m+1)}_{2\nu}$ is proportional
to $J_z+\nu$. As far as we know, this observation is new and we ignore whether
it can be extended to all values of $k$. As a consequence, $P^{(2m)}_{2\nu+1}|j,-\frac{2\nu+1}{2}\rangle=0$
and $P^{(2m+1)}_{2\nu}|j,-\nu\rangle=0$ for all values of $j$. This restricts
the influence of the corresponding crystal-field parameters.




\section{Operator equivalents and parameters for crystal field}
Eq.~(\ref{generalHamiltonian}) is a consequence of the symmetry of the studied system. Hence, it is used as a semi-empirical model to compare to experiments and assign numerical values to the parameters $B^k_q$. However, it is important to note these parameters can also be computed through \textit{ab initio} methods, with programs like ORCA~\cite{chibotaru_ab_2012, ungur_ab_2017}, PyCrystalField~\cite{scheie_pycrystalfield_2021} or SIMPRE~\cite{baldovi_simpre_2013}. The numerical parametrization of the crystal field requires one to adopt a specific normalization of the tensor operators. When investigating the effect of the crystal field on the 4\textit{f} shell of rare earths, neutron scattering literature generally uses Stevens formalism developed on tesseral harmonics operators, whereas x-ray absorption spectroscopy uses Racah spherical tensors operators following Wybourne normalization, i.e. on renormalized spherical harmonics $C^{(k)}_{q}$. However, before going into further details, it must be understood that the encountered usage of crystal-field parameters in the literature has no meaning if it is not along a proper description of $H_\text{CF}$, with its convention and normalization. In his convention, Wybourne proposed to define $H_\text{CF}$ as

\begin{equation}
    H_\text{CF}=\sum_{kq} \hat{B}^k_q  {C}^{(k)}_q,
    \label{eq:HCEFWybourne}
\end{equation}
where the crystal-field parameters $\hat{B}^k_q$ are complex (note the hat emphasizing their complex form) and $(\hat{B}^k_{q})^*=(-1)^q\hat{B}^k_{-q}$. $C^{(k)}_q$ are Wybourne's spherical tensor operators $T^{(k)}_q$ equivalents to the renormalized spherical harmonics defined in Eq.~(\ref{eq:a_kWybourne}) that act on $|JJ_z\rangle$ and give
\begin{align*}
    {C}^{(k)}_q &= \sqrt{\frac{4\pi}{2k+1}}Y_{k,q}.
\end{align*}


On the other side, the Stevens operator equivalents ${O}^q_k(J)$ can be expressed in terms of $J$ within the Coulombian point-charge framework of the crystal-field interaction:

\begin{equation*}
H_\text{CF}(J)= \sum_{kq} A^q_k\langle r^k\rangle\langle J\|\theta_k\|J\rangle{O}^q_k(J),
\end{equation*}
where the real $A^q_k\langle r^k\rangle \langle J\|\theta_k\| J\rangle$ are called the crystal-field parameters. $\langle J\|\theta_k\| J\rangle$ is a multiplicative factor defined for each trivalent rare-earth ion by Stevens in Table 1.1 ref.~\cite{stevens_matrix_1952} or Table VI in ref.~\cite{hutchings_point-charge_1964} (as $\alpha = \theta_2$, $\beta = \theta_4$ and $\gamma = \theta_6$). For the $3d$ elements, such multiplicative factors can be found in Table VII from ref.~\cite{hutchings_point-charge_1964}. In the spectroscopy community, and more particularly in the neutron scattering community, the confusing parameters $B^q_k = A^q_k\langle r^k\rangle\langle J\|\theta_k\|J\rangle$ are usually employed. 

As Hamiltonians need to be Hermitian, Wybourne's definition $H_\text{CF}$ in Eq.~(\ref{eq:HCEFWybourne}) is often preferably expressed in terms of Hermitian operators. Eq.~(\ref{duality}) gives the Hermitian conjugate (or adjoint) of a tensor operator. The Hermitian operators $C^{(k)}_{q}(\text{c},\text{s})$ can thus be defined with the following combinations of $C^{(k)}_{\pm q}$ as:

\begin{subequations}
    \begin{align}
        C^{(k)}_{0}&= C^{(k)}_{0}\\
        C^{(k)}_{q>0}(\text{c})&=\big(C^{(k)}_{-q}+(-1)^qC^{(k)}_{+q}\big),\\
        C^{(k)}_{q>0}(\text{s})&= \text{i}\big(C^{(k)}_{-q}-(-1)^qC^{(k)}_{+q}\big).
    \end{align}
    \label{CHermitianComb}
\end{subequations}
Instead of introducing a new notation for this Hermitian expansion of $H_\text{CF}(L)$, the crystal field is preferentially written in terms of the explicit Hermitian combinations introduced in Eq. (\ref{CHermitianComb}) with real parameters $B^k_{\pm q}$, viz:

\begin{align}
H_\text{CF}&=\sum_k B^k_0C^{(k)}_0 + \sum_{k,q>0}\bigg[B^k_{+q}\big(C^{(k)}_{-q}+(-1)^qC^{(k)}_{+q}\big)+B^k_{-q}\textrm{i}\big(C^{(k)}_{-q}-(-1)^qC^{(k)}_{+q}\big)\bigg].
\label{HcefHermitian}
\end{align}
But one can of course replace the combinations of $C^{(k)}_{\pm q}$ according to Eq.~(\ref{CHermitianComb}) as well as $B^k_{+q}$ with $B^{k}_{q}\text{c}$ and $B^k_{-q}$ with $B^{k}_{q}\text{s}$. The relation of $\hat{B}^k_{q}$ as a function of $B^k_{\pm q}$ also comes as:

\begin{align}
\label{eq:complexB}
\hat{B}^k_{\pm q}= (\mp 1)^q \big(B^k_{+q} \mp \text{i}B^k_{-q}\big).
\end{align}
If a parameter $B^k_{\pm q}$ does not exist for a given $q$, $\hat{B}^k_{\pm q}$ shall then be calculated by considering it as 0. Because of the differential basis used for the two definitions, the real crystal-field parameters $B^k_{\pm q}$ from Wybourne expansion differ from Stevens parameters by a factor $\lambda_{k,q}$, which values are listed in Table \ref{tab:lambdakq}. The values of the table are consistent with previously published tables~\cite{biedenharn_angular_1981, ryabov_generation_1999, rotter_using_2004}. The reader should be careful with Table 6-1 published by Wybourne~\cite[p.~165]{wybourne_spectroscopic_1965} as it contains errors for several parameters $B^k_{q}$.

\begin{table*}[ht]
\centering
\begin{ruledtabular}
\begin{tabular}{c|w{c}{40pt}w{l}{40pt}w{l}{40pt}w{l}{40pt}w{l}{40pt}w{l}{40pt}w{l}{40pt}w{l}{40pt}}
    \diagbox[width=20pt]{$k$}{$|q|$} & $0$ & $1$ & $2$ & $3$ & $4$ & $5$ & $6$ & $7$\\[2ex] \hline
    $0$ & $1$               &                               &                                  & & & & & \\[2ex]
    
    $1$ & $1$               &    $\sqrt{2}$      &                                  & & & & & \\[2ex]
    
    $2$ & $\frac{1}{2}$     & $\sqrt{6}$             & $\frac{1}{2}\sqrt{6}$  & & & & &\\[2ex]
    
    $3$ & $\frac{1}{2}$     &    $\frac{1}{2}\sqrt{3}$       &     $\frac{1}{2}\sqrt{30}$             & $\frac{1}{2}\sqrt{5}$   & & & &\\[2ex]
    
    $4$ & $\frac{1}{8}$     & $\frac{1}{2}\sqrt{5}$  & $\frac{1}{4}\sqrt{10}$   & $\frac{1}{2}\sqrt{35}$ & $\frac{1}{8}\sqrt{70}$  & & &\\[2ex]
    
    $5$ & $\frac{1}{8}$     &    $\frac{1}{8}\sqrt{30}$       &     $\frac{1}{4}\sqrt{210}$             & $\frac{1}{8}\sqrt{35}$  & $\frac{3}{8}\sqrt{70}$  & $\frac{3}{8}\sqrt{7}$ & &\\[2ex]
    
    $6$ & $\frac{1}{16}$    & $\frac{1}{8}\sqrt{42}$ & $\frac{1}{16}\sqrt{105}$ & $\frac{1}{8}\sqrt{105}$ & $\frac{3}{16}\sqrt{14}$ &
    $\frac{3}{8}\sqrt{77}$  & $\frac{1}{16}\sqrt{231}$& \\[2ex]
    
    $7$ & $\frac{1}{16}$    &    $\frac{1}{32}\sqrt{14}$       &     $\frac{1}{16}\sqrt{21}$             & $\frac{1}{32}\sqrt{42}$ & $\frac{1}{16}\sqrt{462}$  & $\frac{1}{32}\sqrt{462}$ & $\frac{1}{16}\sqrt{3003}$ & $\frac{1}{32}\sqrt{858}$\\[2ex]

\end{tabular}
\end{ruledtabular}

\caption{$\lambda_{k,q}$ factors relating Wybourne $B^k_{q}$ parameters to Stevens parameters}
\label{tab:lambdakq}
\end{table*}
The relations between the Wybourne parameters $B^k_{q}$ and the Stevens parameters $B_k^{q}$ and $A_k^{q}\langle r^k\rangle$ acting on ($J$) are the following:

\begin{align}
B^k_{\pm q} = \frac{B_k^{\pm q}(J)}{\lambda_{k,q}\theta_k} = \frac{A_k^{\pm q}\langle r^k\rangle(J)}{\lambda_{k,q}}
\label{eq:Stev_Wyb_relation}
\end{align}

\subsection{Matrix elements of Racah's unit tensor operator}
$H_\text{CF}$ can be developed directly from the Racah's unit tensor operators $u^{(k)}_q$. The operators are defined as $U^{(k)} \equiv \sum^n_{i=1} u^{(k)}$ for a shell filled with $n$ electrons and are the normalized equivalent to $C^{(k)}_q$, where $\langle \ell \| u^{(k)}\|\ell'\rangle = 1$ and

\begin{align*}
    u^{(k)}_q = \frac{C^{(k)}_q}{\langle \ell \| C^{(k)}\|\ell'\rangle}.
\end{align*}

Taking back Wybourne's definition in Eq.~(\ref{eq:HCEFWybourne}), we relate the potential associated to the effect of the crystal field on a configuration $\ell^n$ by~\cite[p.~164]{wybourne_spectroscopic_1965}:

\begin{equation}
    \begin{aligned}
    \langle \ell^n\alpha LSJM_J|H_\text{CF}|\ell^n\alpha'
    L'SJ'M_J'\rangle
     =\sum_{kq}B^k_q &\langle \ell^n\alpha LSJM_J|U^{(k)}_q|\ell^n\alpha'
    L'SJ'M_J'\rangle
    \\\times&\langle \ell\|C^{(k)}\|\ell\rangle
    \end{aligned}
\label{eq:WybourneVpot}
\end{equation}

Because $S$ is invarient in an electric field, $S' = S$. $\langle \ell\|C^{(k)}\|\ell\rangle$ contains the $\ell$-state dependence of the electrons and comes as 

\begin{align*}
\langle \ell \| C^{(k)} \|\ell\rangle &= (-1)^\ell (2\ell+1)
 \threej{\ell}{k}{\ell}{0}{0}{0},
\end{align*}
with~\cite[p.~251]{khersonskii_quantum_1988}
\begin{align*}
\threej{\ell}{k}{\ell}{0}{0}{0} &=  \frac{1}{\sqrt{2k+1}} C^{k0}_{\ell 0\ell 0}
= (-1)^{\ell-k/2} \frac{(\ell+k/2)!}{(k/2)!(k/2)!(\ell-k/2)!} 
\left(\frac{(k!)^2(2\ell-k)!}{(2\ell+k+1)!}\right)^{\sfrac{1}{2}}\
\end{align*}

where the Clebsh-Gordan coefficient $C^{jm'}_{jmkq}$ is not to be mistaken for the operator $C^{(k)}_q$.

\subsubsection{One-electron case}
In the case where the shell $\ell$ is occupied by only 1 electron (or, similarly, if it has only one hole), we can use formula (40) of \cite[p.~481]{khersonskii_quantum_1988}

\begin{align*}
\langle \ell^1 \ell sjm\|U^{(k)}_q\|\ell^1\ell sjm'\rangle &= 
(-1)^{j-m'}\threej{j}{k}{j}{-m'}{q}{m}
(-1)^{j+\ell+s-k} (2j+1)
\sixj{\ell}{s}{j}{j}{k}{\ell} \langle \ell\| C^{(k)}\|\ell\rangle.
\end{align*}

\subsubsection{Many-electron case}
We consider now the case where the shell $\ell$ is occupied by $n$ electrons ($0\le n \le 2(2\ell+1)$). The Wigner-Eckart theorem gives

\begin{equation*}
    \begin{aligned}
    \langle \ell^n\alpha LSJM_J\|U^{(k)}_q\|\ell^n\alpha'L'SJ'M_J'\rangle &= 
    (-1)^{J-M_J}\threej{J}{k}{J'}{-M_J}{q}{M_J'}
    \\&\times\langle \ell^n\alpha LSJ\|U^{(k)}\|\ell^n\alpha'L'SJ'\rangle
    \end{aligned}
\end{equation*}

from Eq.~(\ref{eq:WybourneVpot}), where

\begin{equation*}
    \begin{aligned}
    \langle \ell^n\alpha LSJ\|U^{(k)}\|\ell^n\alpha'L'SJ'\rangle&=(-1)^{S+L+J'+k}\sqrt{(2J+1)(2J'+1)}
    \\&\times\sixj{J}{J'}{k}{L'}{L}{S}\langle \ell^n\alpha LS\|U^{(k)}\|\ell^n\alpha'
    L'S'\rangle
    \end{aligned}
\end{equation*}
We now use Eq.~(11.53) from Cowan~\cite[p.~317]{cowan_theory_1981} to get:

\begin{equation}
    \begin{aligned}
    \langle \ell^n\alpha LS\|U^{(k)}\|\ell^n\alpha'
    L'S'\rangle&=\delta_{SS'}n(-1)^{\ell+L+k} \sqrt{(2L+1)(2L'+1)}\\
    &\times
    \sum_{\noverline{\alpha}\noverline{L}\noverline{S}}
    (-1)^{\noverline{L}}
    \sixj{\ell}{k}{\ell}{L}{\noverline{L}}{L'}
    \big(\ell^n \alpha L S\llbrace\ell^{n-1}\noverline{\alpha}
    \noverline{L}\noverline{S}\big)
    \big(\ell^{n-1}\noverline{\alpha}
    \noverline{L}\noverline{S} \rrbrace \ell^n \alpha' L' S'
    \big).
    \end{aligned}
    \label{manybodyCEF}        
\end{equation}
Note that the 6-j symbol introduced here is equivalent, despite different organization of the quantum numbers, to the 6-j symbol used in Eq.~(2-92) by Wybourne~\cite[p.~32]{wybourne_spectroscopic_1965}.
\subsection{Coefficients of fractional parentage}
The term $\big(\ell^{n-1}\noverline{\alpha}\noverline{L}\noverline{S} \rrbrace \ell^n \alpha' L' S'\big)$ in Eq.~(\ref{manybodyCEF}) is a coefficient called coefficient of fractional parentage (cfp). The associated term $\big(\ell^n \alpha LS \llbrace \ell^{n-1}\noverline{\alpha}\noverline{L}\noverline{S}\big)$ is the inverse of $\big(\ell^{n-1}\noverline{\alpha}\noverline{L}\noverline{S}\rrbrace\ell^n \alpha LS \big)$. Such coefficient allows the expression of an allowed term $|\alpha LS\rangle$ of $\ell^n$, called a \textit{daughter} (or \textit{offspring}), as a function of its \textit{parents} $\noverline{L}\noverline{S}$ of $\ell^{n-1}$ as

\begin{align}
| \ell^n\alpha LS\rangle
 &= \sum_{\noverline{L}\noverline{S}}\big|(\ell^{n-1}\noverline{L}\noverline{S},\ell)LS\rangle \big(\ell^{n-1}
\noverline{L}\noverline{S} \rrbrace \ell^n \alpha' L' S'\big).
\label{linearcomb}
\end{align}
However, these coefficients are difficult to obtain and numerous. The correct description of an unfilled shell (e.g. $f^5$) can necessitate the calculation of a lot of cfp. Fortunately, the system-independent nature of the cfp allows  for their single calculation to serve to many usages later on. The calculation of these cfp has thus been done over time by different authors. A first complete list came from Nielson \& Koster~\cite{nielson_spectroscopic_1963} for the one-body cfp, useful for the calculation of matrix elements of operators representing crystal-field interaction or direct Coulomb repulsion of electrons in a single-shell (often denoted as $F^k_{\ell\ell}$). Later, Donlan~\cite{donlan_twoelectron_1970} computed some of the two-body cfp, useful for calculating matrix elements of operators like the direct and indirect Coulomb repulsion operators of electrons in two different shells (often denoted as $F^k_{\ell_1\ell_2}$ and $G^k_{\ell_1\ell_2}$). Yet, no complete set was given for the two- (and higher-) body cfp. Dobromir D. Velkov computed complete minimal subsets of multi-body cfp for the $p$, $d$ and $f$ electrons in the framework of his PhD thesis~\cite{judd_algebraic_2000,velkov_multi-electron_2000} and are to this date the most complete calculated set we could find. As part of this work, Velkov kindly provided us the PDF file listing his computed tables of minimal subsets of multi-body cfp, in base-36 positional notation. We implemented in Python a numerical extraction of the PDF data and compiled it to a HDF5 file-format database, where the cfp are given directly in the decimal system (base-10) to ease their usage. In fact, the advantages of the base-36 format used by the previous paper-published databases are not relevant in a digital database. The database is available in ref.~\cite{duros_coefficients_2024} and follows Racah and Nielson \& Koster conventions for the listing of the cfp. The cfp are normalized so that

\begin{align*}
\sum_{\noverline{L}\noverline{S}}&=\big(\ell^n\noverline{L}\noverline{S}\rrbrace\ell^{(n+1)}\alpha LS \big)^2 = 1
\end{align*}
to meet requirements of Eq.~(\ref{linearcomb}). 
In the Nielson \& Koster number representation, the cfp are presented in a power-of-primes writing scheme, consisting of thirteen primes $a_0$, $a_1$, $a_2$, $a_3$, $a_4$, $a_5$, $a_6$, $a_7$, $a_8$, $a_9$, $a_{10}$, $a_{11}$ and $a_{12}$, such that

\begin{equation*}
    \text{cfp} = a_0 \times \bigg(\prod^{12}_{i=1}p_i^{a_i}\bigg)^{\frac{1}{2}}
\end{equation*}
where $p_i$ is the $\text{i}^\text{th}$ prime among 2, 3, 5, 7, 11, 13, 17, 19, 23, 29, 31 and 37. Because more than one multiplet of a given $\mathit{LS}$ pair may occur, Racah differentiated them by introducing additional quantum numbers, not of physical significance, by reference to the grand parents of the multiplet and to the properties of mathematical groups $\text{R}_5$ for the $d^n$ configurations and $\text{R}_7$ and $\text{G}_2$ for the $f^n$ configurations. These additional numbers are respectively called seniority number, $\text{R}_5$ representation symbol and $\text{R}_7$ and $\text{G}_2$ representation symbols. To avoid cumbersome listing of the states, the several $\mathit{LS}$-sharing terms within a same configuration are listed in the database with a sequential index, e.g. 2D1 and 2D2. The correspondence between the aforementioned symbols and the sequential index for a given $\mathit{LS}$ pair is available in the ``states" database group. This index actually corresponds to $\alpha$ in Eq.~(\ref{manybodyCEF}). A simple Python 3 script is provided to compute the matrix elements of $U^{(k)}_q$ using the HDF5 database.

\subsection{Parametrization of the crystal field in numerical tools}
Several numerical tools with an implementation of crystal field for rare-earth compounds appeared over time, either to fit different type of experiments (INS, RIXS), either to use point charge model to determine crystal-field parameters. Unfortunately, these different tools came with as many implementations of the crystal field as we can find in the literature. Some of them have already been reviewed and only a few will be mentioned here. We try to strictly respect the writing in use in each tools to ease the differentiation with the parameters we introduced earlier.

\subsubsection{McPhase}
McPhase~\cite{rotter_using_2004} is an open-source program that calculates several magnetic properties of materials. Its module \texttt{so1ion} uses different types of parameters (and thus, expansions) for the expression of the crystal field acting on the ion of interest. Those parameters are listed in Table 5, appendix ``Crystal Field and Parameter Conventions" from its own manual. McPhase uses the letters $l$ and $m$ for the rank and the quantum magnetic number that we describe with the letters $k$ and $q$, respectively. Also, note that in the documentation, the positions of $l$ and $m$ are the same in the two conventions from Stevens and Wybourne, which is not what we observe usually in the literature. McPhase $B_{l}^m$ and $A_{lm}$ correspond then to Stevens $B^q_k$ and $A^q_k$. For the parameters following the Wybourne normalization, $D_{l}^m$ corresponds to $\hat{B}^k_q$ introduced in Eq.~\ref{eq:HCEFWybourne}. We however bring the parameter $L_{l}^m$ to the users attention. According to the software-provided examples, $L_{l}^m$ has to be given in the \textbf{input} files as \texttt{Llm} for $m>0$ and \texttt{LlmS} for $m<0$. However, the \textbf{output} files lists the \textit{same} parameters $L_{l}^m$ as \texttt{Llm} for $m>0$ and \texttt{Ll-m} for $m<0$, with \texttt{Ll-m} being the opposite of the input \texttt{LlmS}, i.e. $\texttt{Ll-m} = -\texttt{LlmS}$. We thus relate our real $B^k_q$ from Eq.~\ref{HcefHermitian} to what can be described in McPhase input files by \texttt{LlmS}. 

\subsubsection{PyCrystalField}\label{sec:PCF}
PyCrystalField is an open-source Python package designed to compute the $H_\text{CF}$ of single-ions. To do so, the package uses the electrostatic PCM briefly introduced in section~\ref{sec:intro} by extracting the crystallographic data from a CIF file. The calculation returns a list of crystal-field parameters expressed in the Stevens formalism. PyCrystalField uses the letters $n$ and $m$ for the rank and the quantum magnetic number that we describe respectively with the letters $k$ and $q$. PyCrystalField $B^m_n$ (\texttt{B\_n m} in the code) correspond to the real Stevens $B^q_k$. An interesting advantage of the package is that $H_\text{CF}$ can be developed in the Stevens formalism on the $\mathit{LS}$ basis. If a spin-orbit interaction parameter \texttt{LS\_Coupling} is given to the function, the $B^m_n$ returned in the output are scaled to the $\mathit{LS}$ basis as followed:

\begin{align}
    B_n^{\pm m} (L)& = \frac{B_n^{\pm m}(J)}{\theta_{k=n}}\times\gamma_{k=n}
    \label{eq:PCFdef}
\end{align}
(note the position of the indices for $B_n^{\pm m}$ denoting Stevens parameters). The relation between the Wybourne and PyCrystalField $B_n^{\pm m} (L)$ parameters comes thus easily as
\begin{align}
    B^k_{\pm q}& = \frac{B_{n}^{\pm {m}} (L)}{\lambda_{kq}\gamma_{k}}.
    \label{eq:PCFrel}
\end{align}
PyCrystalField calculated the values of $\gamma_{k}$ which are listed in their source code, accessible on GitHub. The method to derive these values is detailed in ref.~\cite{scheie_pycrystalfield_2021}.

\subsubsection{SIMPRE}
On a analogous concept to PyCrystalField, SIMPRE is an open-source Fortran77 code using the electrostatic PCM to predict the magnetic properties of rare-earth compounds. It gives as an output two Stevens crystal-field parameters: $A^q_k\langle r^k\rangle$ and $B^q_k$ that are the Stevens parameters we define in the same fashion and are related to Wybourne parameters following Eq.~(\ref{eq:Stev_Wyb_relation}).

\subsubsection{SPECTRE}
SPECTRE~\cite{boothroyd_spectre_1990} is a closed-source program developed for calculating spectroscopic and magnetic single-ion properties of rare-earth ions in crystals. The calculation employs intermediate coupling basis states, and allows mixing of all terms within the $f^n$ configuration. The program thus uses a description of the crystal field on Racah spherical operators, as defined in Eq.~(\ref{HcefHermitian}). Parameters $B^k_{q}\text{(c)}$ and $B^k_{q}\text{(s)}$ are the same as the $B^k_{q}(\text{c})$ and $B^k_{q}(\text{s})$ used in the present paper.

\subsubsection{Quanty}
Quanty is a closed-source many-body script language based on Lua which allows calculation of quantum mechanical problems. In Quanty, the crystal field is expanded on Racah spherical operators following Wybourne normalization. Quanty uses the letters $k$ and $m$ for the rank and the quantum magnetic number that we describe with the letters $k$ and $q$, respectively. The documentation mentions the following: ``one can create crystal-field operators with the function \texttt{NewOperator()} with, as a first input, the string \texttt{'CF'}. The function furthermore needs to know which set of $k$ and $m$ the operator will be expanded on, and the effective potential needed to be expanded on the operator, i.e. the crystal-field parameter. This is given as a list of the form $\{\{k_1,m_1,A_{k_1,m_1}\},\{k_2,m_2,A_{k_2,m_2}\},\dots\}$". The choice of $A_{k,m}$ for this variable name can be quite confusing for the user as it reminds the Stevens $A_{k}^q$, while it is actually the complex parameters $\hat{B}^k_q$. In order to express correctly the aforementioned effective potential for a given set of $k$ and $q$ ($m$ in Quanty), we can see how Eq. (\ref{HcefHermitian}) can be re-written as a combination of parameters $B^k_{\pm q}$ with respect to the operators $C^{(k)}_{\pm q}$:

\begin{align*}
    H_\text{CF}&=\sum_k B^k_0C^{(k)}_0 + \sum_{k,q>0}\bigg[C^{(k)}_{+q}\big((-1)^qB^k_{+q}-(-1)^q\text{i}B^k_{-q}\big)+C^{(k)}_{-q}\big(B^k_{+q}+\text{i}B^k_{-q}\big)\bigg]
    \\&\equiv\sum_k A_{k,0}C^{(k)}_0 + \sum_{k,m>0}\big(C^{(k)}_{+m}A_{k,+m}+C^{(k)}_{-m}A_{k,-m}\big)
    \label{HcefHermitian_withB}
\end{align*}
where the parameters $A_{k,\pm m} \equiv \hat{B}^k_q$ from Eq.~\ref{eq:complexB}.

\subsubsection{EDRIXS}

EDRIXS~\cite{wang_edrixs_2019} is an open-source toolkit for simulating XAS and RIXS spectra in a Python environment, based on exact diagonalization of Hamiltonians. Similarly as seen in section~\ref{sec:PCF} with PyCrystalField, one can define the crystal field in EDRIXS on the $\mathit{LS}$ basis by developing Stevens operators as linear combinations of $L_z$ and $L_\pm$. The crystal-field parameters $B_k^q(L)$ follow then the same relation to the real Stevens parameters as seen in Eq.~(\ref{eq:PCFdef}) and to the real Wybourne parameters as seen in Eq.~(\ref{eq:PCFrel}).

\section{Conclusion}
In this paper, we reviewed the existing expressions for the spherical tensor form of Stevens or spin operator equivalents. We gave explicit proofs of these forms, which were not available in the literature and we devised a way to transform expressions in terms of $j$ and $m$ into expressions in terms of $J^2$ and $J_z$. Along the way, we discovered factorization properties of some of the spherical tensors which seem to be new. We hope that the present paper will enable the physics and chemistry communities to avoid having recourse to cumbersome (and generally not fully correct) tables.
In the last part, we give some insights on how crystal-field parameters from different conventions are used in recent numerical tools and we present the relations between them. Eventually, we provide a complete minimal subsets of multi-body fractional parentage coefficients for the $p$, $d$ and $f$ electrons, essential for the calculation of the unit tensor operator $u^{(k)}_q$ that can be used to expand, for example, the crystal-field potential. 
Among the perspectives of this work, we might mention the question whether or not the factorization properties extend to general values of $k$. We also want to mention that some analytical expressions were given for non-diagonal tensor-equivalent operators~\cite{grenet_operator_1978,caola_operator_1979, sadovskii_standard_1991}, i.e. for the case of matrix elements between different values of $j$. However, the generalization of our results to the non-diagonal case seems non trivial. Finally, the usage of the cfp database provided here for the expression of crystal-field operators can be extended to many useful operators.

\begin{acknowledgements}
We are very grateful to Genviève Grenet for pointing out to us that the main formula of her article with Kibler is proved in her PhD thesis. We are also grateful to the Online Encyclopedia of Integer Sequences (https://oeis.org) and to Marjorie Bicknell Johnson whose paper with Hoggatt~\cite{hoggatt_catalan_1976} lead us to Riordan's theorem. Last, we would also like to express our sincere thanks to Dobromir D. Velkov for providing us his full PhD manuscript, including appendixes with the computed list of many-body coefficients of fractional parentage.
\end{acknowledgements}

\clearpage
\end{document}